\documentclass[twocolumn,prb,showpacs,showkeys,floatfix]{revtex4}

\usepackage{graphicx}
\usepackage{dcolumn}
\usepackage{bm}

\begin{document}

\preprint{APS/123-QED}

\title{Temperature dependence of the coercive field in single-domain particle systems}

\author{W. C. Nunes, W. S. D. Folly, J. P. Sinnecker and M. A. Novak}
 \author{}%

\affiliation{Instituto de F\'isica, Universidade Federal do Rio de Janeiro, C. P. 68528, Rio de Janeiro, RJ 21945-970, Brazil}%


\date{\today}

\begin{abstract}
The magnetic properties of Cu$_{97}$Co$_{3}$ and
Cu$_{90}$Co$_{10}$ granular alloys were measured over a wide
temperature range ($2$ to $300$~K). The measurements show an
unusual temperature dependence of the coercive field. A
generalized model is proposed and explains well the experimental
behavior over a wide temperature range. The coexistence of blocked
and unblocked particles for a given temperature rises difficulties
that are solved here by introducing a temperature dependent
blocking temperature. An empirical factor gamma ($\gamma$) arise
from the model and is directly related to the particle
interactions. The proposed generalized model describes well the
experimental results and can be applied to other single-domain
particle system.
\end{abstract}



\maketitle

\section{\label{sec:level1}Introduction}

Since N\'eel's\cite{Neel} pioneering work, the magnetic properties
of single-domain particles have received considerable
attention,\cite{Fiorani} with both technological and academic
motivations. A complete understanding of the magnetic properties
of nanoscopic systems is not simple, in particular because of the
complexity of real nanoparticle assemblies, involving particle
size distributions, magnetic interparticle interactions, and
magnetic anisotropy. Despite these difficulties, many theoretical
and experimental investigations of such systems have provided
important contributions to the understanding of the magnetism in
granular systems.\cite{Fiorani}

An important contribution to the understanding of the magnetic
behavior of nanoparticles was given by Bean and
Livingston\cite{Bean} assuming an assembly of noninteracting
single-domain particles with uniaxial anisotropy. This study was
based on the N\'eel relaxation time
\begin{equation}
\tau=\tau_{0}e^{\frac{KV}{k_{B}T}}, \label{Arrhenius}
\end{equation}
where the characteristic time constant $\tau_{0}$ is normally
taken in the range $ 10^{-11}-10^{-9}$~s, $k_{B}$ is the Boltzmann
constant, $K$ the uniaxial anisotropy constant, and $V$ the
particle volume. $KV$ represents the energy barrier between two
easy directions. According to Bean and Livingston, at a given
observation time ($\tau_{obs}$) there is a critical temperature,
called the blocking temperature $T_{B}$, given by
\begin{equation}
T_{B}=\frac{KV}{\ln\left(\frac{\tau_{obs}}{\tau_{0}}\right)k_{B}},
\label{tb}
\end{equation}
above which the magnetization reversal of an assembly of identical
(same volume) single-domain particles goes from blocked (having
hysteresis) to superparamagnetic-type behavior.

The time and temperature behavior of the magnetic state of such
assembly is understood on the basis of N\'eel relaxation and the
Bean-Livingston criterion (Eqs.~(\ref{Arrhenius}) and (\ref{tb})).
Within this framework the coercive field is expected to decay with
the square root of temperature, reaching zero in thermal
equilibrium.\cite{Bean} As the effects of size distribution and
interparticle interactions cannot be easily included in this
model, this thermal dependence of $H_{C}$ is widely used. The
discrepancies between theory and experiment are usually attributed
to the presence of interparticle interactions and a distribution
of particle sizes.\cite{Fonseca, Batlle}

The effect of dipolar interactions on the coercive field and
remanence of a monodisperse assembly has been recently studied by
Kechakros and Trohidou\cite{Kechakros} using a Monte Carlo
simulation. These authors show that the dipolar interaction slows
down the decay of the remanence and coercive field with
temperature. The ferromagnetic characteristic (hysteresis)
persists at temperatures higher than the blocking temperature
$T_{B}$. However, the existence of a size distribution in real
systems jeopardizes the observations of such behavior in
experimental data.

Studies considering size distribution have been limited to the
contribution of superparamagnetic particles, whose relative
fraction increases with temperature. Although this contribution
has been recognized to be an important factor affecting the
coercive field, a complete description of this effect is still
open.\cite{Kneller, Pfeiffer, Vavassori} As an example, let us
consider Cu-Co granular systems that presents a narrow hysteresis
loop, with a coercive field smaller than $1000$ Oe even at low
temperatures. The isothermal magnetization curve suggests a
superparamagnetic behavior, while the coercive field has an
unusual temperature dependence. This behavior has been
qualitatively attributed to interactions\cite{Nunes} or size
distribution effects.\cite{Ferrarin} Nevertheless, this $H_{C}(T)$
cannot be quantitatively explained by the current models in the
whole temperature range.

The combined effects of size distribution and interparticle
interactions on the magnetic behavior is very complex, because
both affect the energy barriers of the system. Due to this
complexity, the size distributions of such systems obtained by
different magnetic methods\cite{Peleg, Chantrell, Folly, Kliava}
include both effects and may be considered as an `effective size
distributions'.

In this work we develop a generalized method to describe the
temperature dependence of the coercive field of single-domain
particle systems. We shall focus on the specific samples
Cu$_{97}$Co$_{3}$ and Cu$_{90}$Co$_{10}$ granular
ribbons.\cite{Allia, Wecker, Allian} The novelty in this approach
consists in the use of a proper mean blocking temperature, which
is temperature dependent. The temperature behavior of the coercive
field of the studied systems has been successfully described in a
wide temperature range and may be applied to other systems.

\section{The Generalized Model}

Considering an assembly of noninteracting single-domain particles
with uniaxial anisotropy and particle volume $V$, the coercive
field in the temperature range from 0 to $T_{B}$, i.e., when all
particles remain blocked, follows the well-known relation:
\begin{equation}
H_{C}=\alpha
\frac{2K}{M_{S}}\biggl[1-\biggl(\frac{T}{T_{B}}\biggr)^\frac{1}{2}\biggr]
\label{Hc}
\end{equation}
Here $M_{S}$ is the saturation magnetization and $\alpha = 1$ if
the particle easy axes are aligned or $\alpha = 0.48$ if randomly
oriented.\cite{Bean, Stoner}

Some difficulties arise when $H_{C}(T)$ is calculated on a system
with several particle sizes: $(i)$ the distribution of sizes
yields a corresponding distribution of $T_{B}$; $(ii)$ the
magnetization of superparamagnetic particles, whose relative
fraction increases with temperature, makes the average coercive
field smaller than those for the set of particles that remain
blocked. These two points will be discussed in the following
sections.

\subsection{Influence of size distribution}

In granular systems there will be a particle size distribution
which gives rise to a distribution of blocking temperatures TB.
One can

The conventional way used to take into account the particle size
distribution on the coercive field is to define a mean blocking
temperature $\left<T_{B}\right>$ corresponding to a mean volume
$\left<V_{m}\right>$ and then consider $H_{C}(T)$ in
Eq.~(\ref{Hc}) with $T_{B}$ replaced by the average blocking
temperature of all the particles \cite{Chantrell, elhilo, mblanco}

\begin{equation}
\left<T_{B}\right>=\frac{\int_{0}^{\infty}T_{B}f(T_{B})dT_{B}}{\int_{0}^{\infty}f(T_{B})dT_{B}},
\label{tbme}
\end{equation}
where $f(T_{B})$ is the distribution of blocking temperatures.

This artifice may lead to a bad agreement with the experimental
results because it does not take into account point $(i)$
conveniently and neglects point $(ii)$ of the above discussion.
Often, a reasonable agreement with experimental data using this
approach is obtained only for $T$ well below
$\left<T_{B}\right>$,\cite{Fonseca, Batlle} \textit{i.e.}, when
there is a great fraction of blocked particles. In this work, the
assumption $(i)$ and $(ii)$ are used to determine a more realistic
average blocking temperature.

The idea consist in taking into account only the blocked particles
at a given temperature ($T$), i.e., only particles with $T_{B}>T$.
Thus, Eq.~(\ref{Hc}) should be rewritten as
\begin{equation}
H_{CB}=\alpha
\frac{2K}{M_{S}}\biggl[1-\biggl(\frac{T}{\left<T_{B}\right>_{T}}\biggr)^\frac{1}{2}\biggr],
\label{hcbt}
\end{equation}
where $\left<T_{B}\right>_{T}$ is the average blocking
temperature, defined by:
 \begin{equation}
\left<T_{B}\right>_{T}=\frac{\int_{T}^{\infty}T_{B}f(T_{B})dT_{B}}{\int_{T}^{\infty}f(T_{B})dT_{B}},
\label{tbt}
\end{equation}
where the new limits of integration takes into account only the
blocked particles.

\subsection{Influence of superparamagnetic particles}

\begin{figure}[htbp]
\includegraphics[width=6.0cm]{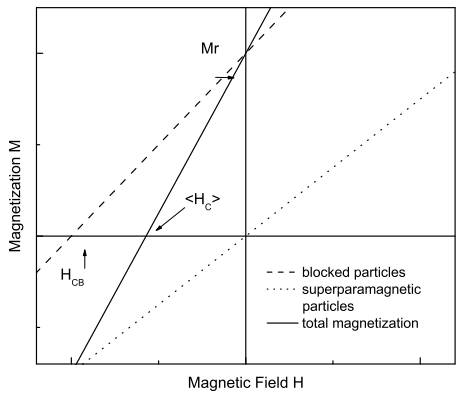}
\caption{\label{epsart} Contribution of the superparamagnetic
particles to the coercive field.}
\end{figure}

The distribution of particle sizes causes a temperature dependence
of the coexistence of both (1) blocked and (2) superparamagnetic
particles. The influence of superparamagnetic particles on the
coercive field was explicitly taken into account by Kneller and
Luborsky.\cite{Kneller} They considered that the magnetization
curves of components (1) and (2) are linear for $H < H_{CB}$ (see
Fig.~\ref{epsart}). Hence $M_{1}=Mr + (Mr/H_{CB})H$ for component
(1) and $M_{2}=\chi_{S} H$ for component (2), where $Mr$ is the
remanence, $H$ the applied magnetic field and $\chi_{S}$ the
superparamagnetic susceptibility. These two components may be
linearly superposed and the average coercive field becomes

\begin{equation}
\left<H_{C}\right>_{T}=\frac{M_{r}(T)}{\chi
_{S}(T)+\frac{M_{r}(T)}{H_{CB}(T)}}. \label{hcm}
\end{equation}

\subsection{Determination of $\left<H_{C}\right>_{T}$}

In order to obtain $\left<H_{C}\right>_{T}$ by Eq.~(\ref{hcm}),
three terms must be evaluated: $M_{r}(T)$, $\chi _{S}(T)$, and
$H_{CB}(T)$ all determined from experiments. For the last two
terms, a good determination of $f(T_{B})$ is needed. The
isothermal remanent magnetization is related to $f(T_{B})$
according to:\cite{Chantrell}
\begin{equation}
I_{r}(T)=\alpha M_{S}\int_{T}^{\infty}f(T_{B})dT_{B} \label{mr}
\end{equation}
Clearly the derivative of Eq.~(\ref{mr}) is a direct measure of
the blocking temperature distribution $f(T_{B})$ i.e.,
$dI_{r}/dT\propto f(T_{B})$.

The superparamagnetic susceptibility has two contributions:
isolated and groups of a few Co atoms, $\chi_{SA}(T)$, and
particles, $\chi_{SP}(T)$. The later is the initial susceptibility
in the low field limit given by $M_{S}^{2}V/3k_{B}T$. For a system
with nonuniform particle sizes, it can be calculated as:
\begin{eqnarray}
\chi _{SP}(T) & = &
\frac{M^{2}_{S}}{3k_{B}T}\int_{0}^{Vc(T)}{Vf(V)dV} \nonumber\\
& = & \frac{25M^{2}_{S}}{3KT}\int_{0}^{T}T_{B}f(T_{B})dT_{B},
\label{xsp}
\end{eqnarray}
where the linear relation between $T_{B}$ and $V$ (Eq.~(\ref{tb}))
was used to express $\chi_{SP}(T)$ in terms of $T_{B}$, and the
critical volume $V_{c}$ above which the particle is blocked. Thus,
the superparamagnetic susceptibility can be written as
\begin{equation}
\chi
_{S}(T)=\frac{25M^{2}_{S}}{3KT}\int_{0}^{T}T_{B}f(T_{B})dT_{B}+\frac{C}{T},
\label{xs}
\end{equation}
where the second term is the contribution of groups of a few Co atoms
defined in terms of the Curie constant $C$. Finally $H_{CB}(T)$ is
determined from the Eqs. (\ref{hcbt}) and (\ref{tbt}).

Alternatively, $f(T_{B})$ can also be determined from zero field
cooled (ZFC) and field cooled (FC) magnetization
experiments\cite{Peleg} when one is sure that there are no blocked
particles at the highest measuring temperature, which is not
always the case.

\section{Experimental}

We have studied two Cu-Co ribbons of nominal composition
Cu$_{97}$Co$_{3}$ and Cu$_{90}$Co$_{10}$ in the as cast form.
These samples were prepared by melt spinning as described in
reference~\onlinecite{Allia}.

The hysteresis loops $M(H)$ of these samples were measured in the
temperature range from $2$ to $300$~K and maximum field of
$90$~kOe.

The temperature dependence of the coercive field, and remanent
magnetization were determined from the hysteresis loops. The
saturation magnetization was determined by extrapolation of
$M(1/H)$ for $1/H = 0$, at $2$~K.

ZFC curves were measured cooling the system in zero magnetic field
and measuring during warming with an external field applied. The
FC curves were measured during the cooling procedure with field.

The $I_{r}(T)$ curve was measured cooling the systems in zero
magnetic field from room temperature. At each temperature the
samples were submitted to a field of $7$~kOe which gives a
negligible remanent field in the superconducting magnet and is
well above the field where the hysteresis loop closes. Following
the same procedure used by Chantrell \textit{et
al}.\cite{Chantrell}, $I_{r}(T)$ was determined by waiting $100$~s
after the field is set to zero.

All measurements were performed using a Quantum Design Physical
Property Measurement System (PPMS) model 6000.

\section{Results and Discussion}

\subsection{Distribution of energy barriers and system nanostructure}

The morphology of a granular system consisting of magnetic
precipitates plays an important role on the macroscopic behavior.
Many efforts were made to investigate the morphology of granules
in Cu-Co alloys. It is very difficult to use electron microscopy
to identify Co granules in a Cu matrix due to the almost identical
atomic scattering factor of Cu and Co atoms and very similar
lattice parameter between the Cu matrix and Co
granules~\cite{hutten}. Other experimental techniques could only
be used to estimate the average size distribution
\cite{yu,Lopez,GarciaPrieto}.

\begin{figure}[htbp]
\includegraphics[width=6.0cm]{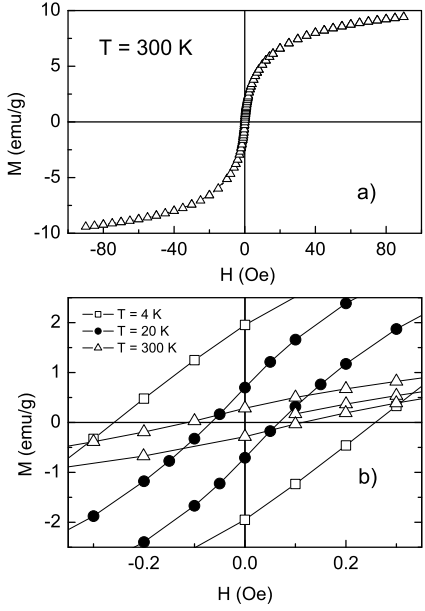}
\caption{\label{mh}(a) Hysteresis loop for the Cu$_{90}$Co$_{10}$
as cast sample at room temperature. (b) A detail of the narrow
hysteresis at different temperatures.}
\end{figure}

In many nanoparticle systems the experimental magnetization curves
are remarkably well fitted using a superposition of properly
weighted Langevin curves, usually considering a log-normal
distribution.\cite{Ferrari} Making the assumptions that the
magnetization of each spherical particle is independent of its
volume, the particle size distribution may be obtained. In fact,
the $M(H)$ curves for the sample Cu$_{90}$Co$_{10}$ shown in
Fig.~\ref{mh}(a) exhibits a superparamagnetic shape, the same
occurring also for Cu$_{97}$Co$_{3}$.

However, $M(H)$ is more sensitive to the average magnetic moment
(or average particle size) than the width or distribution
shape.\cite{Luis} For this reason the use $M(H)$ to determine
$f(T_{B})$ is not a good choice. In addition, a closer look to the
small hysteresis shows that while the remanence decays with
temperature, the coercive field at $20$~K is smaller than at $4$
and $300$~K (see Fig.~\ref{mh}(b)). This issue will be discussed
in next section.

\begin{figure}[htbp]
\includegraphics[width=6.0cm]{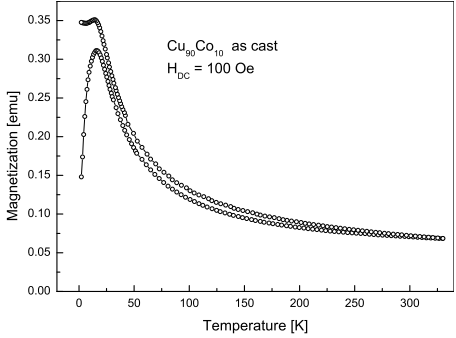}
\caption{\label{zfcfc}Zero field cooled and field cooled curves
for the Cu$_{90}$Co$_{10}$ as cast measured at $H_{DC}$=100~Oe up
to $T=330$~K.}
\end{figure}

We first used ZFC/FC curves (shown in Fig.~\ref{zfcfc}) as usually
done to determine the distribution $f(T_{B})$. Is is clear that
the warming and cooling curves do not close as to the highest
measured temperature. The histeresis in the $M(H)$ curve (see
fig.~\ref{mh}~b) at $300$~K show the presence of blocked
particles. The determination of the distribution $f(T_{B})$ using
this ZFC/FC curves would clearly lead to an inaccurate result
because it would neglect the remaining blocked particles above
$T=330$~K. We found thus better to determine the distribution by
the use of eq.~\ref{mr} and teh experimental remanence.

\begin{figure}[htbp]
\includegraphics[width=6.0cm]{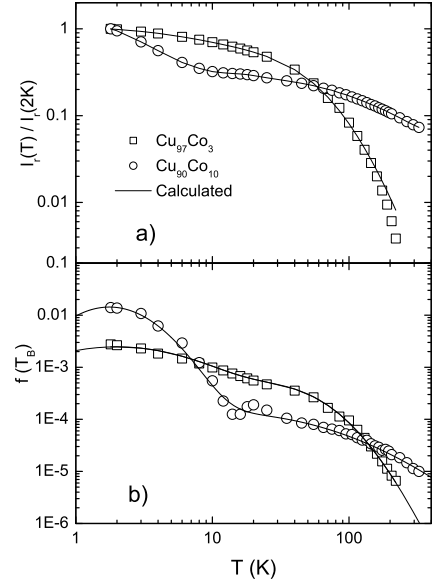}
\caption{\label{distrib}(a) Remanent magnetization for
Cu$_{97}$Co$_{3}$ (square symbols) and Cu$_{90}$Co$_{10}$
(circles). (b) Distribution calculated using Eqs.~(\ref{mr}) and
(\ref{ftb}).}
\end{figure}

Figure \ref{distrib}(a) shows the temperature dependence of the
isothermal remanence of Cu$_{97}$Co$_{3}$ and Cu$_{90}$Co$_{10}$
samples. Cu$_{90}$Co$_{10}$ presents two inflexion points
suggesting two mean blocking temperatures ($\left<T_{B1}\right>$
and $\left<T_{B2}\right>$). This leads us to assume $f(T_{B})$ as
being the sum of two log-normal distributions:

\begin{eqnarray}
f(T_{B}) & = & \frac{A}{T_{B}\sqrt{2\pi } \sigma _{1}}\exp\biggl{[}-\frac{1}{2\sigma ^2_{1}}\ln^{2}\biggl{(}\frac{T_{B}}{\left<T_{B1}\right>}\biggr{)}\biggr{]} \nonumber\\
       &   & + \frac{1-A}{T_{B}\sqrt{2\pi} \sigma _{2}}\exp\biggl{[}{-\frac{1}{2\sigma^2_{2}}\ln^{2}{\biggl{(}\frac{T_{B}}{\left<T_{B2}\right>}}}\biggr{)}\biggr{]} \label{ftb}
\end{eqnarray}

The lines in Fig.~\ref{distrib} were obtained fitting the
experimental data to the integrated Eq.~(\ref{mr}) using the above
$f(T_{B})$ distribution. The free parameters used in the fit were
$\sigma_{1}$, $\left<T_{B1}\right>$, $\sigma_{2}$,
$\left<T_{B2}\right>$ and the weighting factor $A$, all shown in
Table I. Since $KV\propto T_{B}$, $f(T_{B})$ represents the
distribution of energy barriers.

We can see in Fig.~\ref{distrib}(a) that the agreement between the
theoretical and experimental curves is good for both samples.
Figure \ref{distrib}(b) shows the distribution of energy barriers
obtained from the derivative of the isothermal remanence decay
(Eq.~(\ref{mr})). The obtained $f(T_{B})$ confirm the
inhomogeneous  magnetic nanostructure observed by other
authors.\cite{Panissod, GarciaPrieto, Lopez, Wang}

\begin{table}[hp] \centering%
\caption{\label{pftb} Blocking temperature distribution
parameters}
\begin{tabular}{c c c c c c c}
\hline \hline
Sample      & $\sigma_{1}$  & $<T_{B1}>$ (K)& $\sigma_{2}$  & $<T_{B2}> (K)$& $A (\%$) \\
\hline
Co$_{3}$Cu$_{97}$   &1.2    &8.4    &0.6    &53 &61\\
Co$_{10}$Cu$_{90}$ &0.7 &2.7    &1.4    &128 &78\\
\hline \hline
\end{tabular}
\end{table}

Some characteristics of the nanostructure can be inferred from
above results. For Cu$_{90}$Co$_{10}$ there is a large number of
small particles (small energy barriers) responsible for the low
temperature maximum, and a few large particles responsible for
second peak in the energy barrier (see Fig.~\ref{distrib}(b)). For
the less concentrated sample (Cu$_{97}$Co$_{3}$), the energy
barrier of the two groups of particles (small and large) is
expected to be smaller than the observed in more concentrated
sample, owing to the corresponding reduction in the particle
sizes. However, we observe that $\left<T_{B1}\right>$ is bigger
than for the more concentrated sample (see Table \ref{pftb}). Such
behavior can be related to the higher surface anisotropy as the
particle size decreases.\cite{Luis}

\subsection{ Coercive field}

\begin{figure}[htbp]
\includegraphics[width=6.0cm]{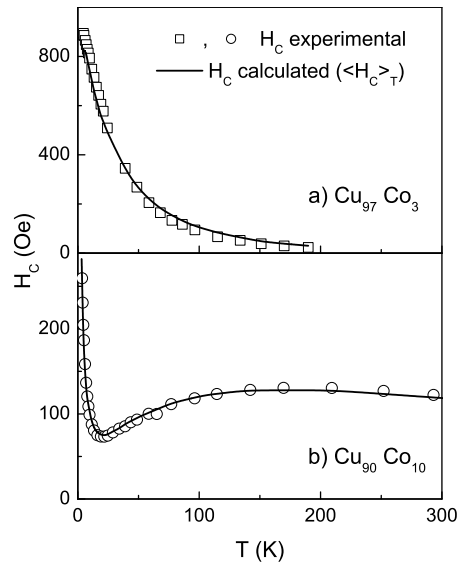}
\caption{\label{hcf}Coercive field $H_{C}$ vs. temperature for the
two samples investigated: experimental (symbols) and calculated
with the generalized model (line).}
\end{figure}

The temperature dependence of the coercive field $H_{C}(T)$ is
shown in Fig.~\ref{hcf}. While the more diluted sample presents
the usual decrease with $T$, the Cu$_{90}$Co$_{10}$ sample
exhibits an unusual $H_{C}(T)$ with a sharp decrease up to $20$~K,
followed by an increase and a maximum around $180$~K.

\begin{figure}[hb]
\includegraphics[width=6.0cm]{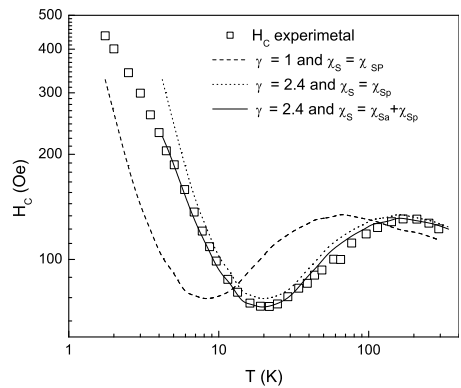}
\caption{\label{hcsteps} $H_{C}$ calculations and experimental
data for Cu$_{90}$Co$_{10}$. The full line is the
$\left<H_{C}\right>_{T}$ obtained considering interaction and
groups of a few Co atoms.}
\end{figure}

The solid lines were calculated by use of Eq.~(\ref{hcm}) with
$f(T_{B})$ obtained previously and adjusting the anisotropy
constant $K$ in Eqs.~(\ref{hcbt}) and (\ref{xs}). Some additional
considerations has to be made to obtain a good agreement with the
data. The straight calculation of $H_{C}(T)$ gives the dashed line
shown in Fig.~\ref{hcsteps}. It is clear that a better agreement
could be obtained by a horizontal shift in this plot. It is well
known that barrier distributions $f(T_{B})$ are field
dependent.\cite{Peleg, Denardin} The distribution of energy
barriers obtained from magnetization measurements shows a
temperature shift, which is a function of either the applied field
or the magnetization state of the sample.\cite{Peleg, Garciadel}
Allia \textit{et al}.\cite{Allian} have proposed that the effect
of interparticle interactions can be pictured by the use of an
additional temperature term in the Langevin function. In our case
the results suggest to use $f(\gamma T_{B})$ instead of
$f(T_{B})$, were $\gamma$ is an empirical parameter that takes
into account the effect of random interactions.  This leads to a
better agreement, as shown by the dotted line curve. So far we did
not take into account the Curie term in Eq.~(\ref{xs}). By doing
so, the agreement with experimental data becomes excellent as
shown by the solid lines in Figs.~\ref{hcf} and \ref{hcsteps} (for
both samples). The best obtained parameters are shown in Table II.

\begin{table}[hp] \centering%
\caption{\label{hctab} Best adjusted parameters }
\begin{tabular}{c c c c}
\hline \hline
Sample                      & $K$ (erg/cm$^{3}$)  & $C$ (emu*K/Oe*cm$^{3}$) & $\gamma $ \\
\hline
Co$_{3}$Cu$_{97}$ as cast         & $5.0*10^{6}$  & $2.3$       & $1.4$ \\
Co$_{10}$Cu$_{90}$ as cast        & $3.2*10^{6}$  & $1.4$       & $2.4$ \\
\hline \hline
\end{tabular}
\end{table}

We can see in the Table II that $K$ is smaller for the more
concentred sample, which is consistent with the relative reduction
of the surface anisotropy contribution with the size of the
nanoparticles.\cite{Respaud, Luis} Note that the Co$_{3}$Cu$_{97}$
sample presents a higher concentration of isolated groups of a few
Co atoms and smaller interaction parameter $\gamma $. The expected
stronger interaction (and $\gamma $) in the Co$_{10}$Cu$_{90}$
sample may also imply in a reduction of $K$, as suggested by the
random anisotropy model.\cite{Nunes} The application of this model
to a system with a negligible interaction provides a good
description of Hc(T) with a gamma temperature shift in the
distribution equal to 1 ($ \gamma = 1 $).

\begin{figure}[htbp]
\includegraphics[width=6.0cm]{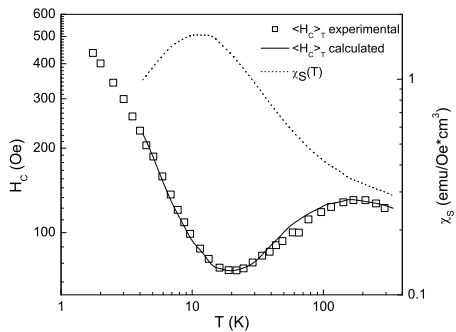}
\caption{\label{hcxs}Calculated $\chi_{S}$ (dotted line),
$\left<H_{C}\right>_{T}$ (solid line), and experimental data (open
symbols) vs. temperature of the Cu$_{90}$Co$_{10}$ sample. }
\end{figure}

The interesting behavior of $H_{C}(T)$ of the Cu$_{90}$Co$_{10}$
sample can be understood in terms of $\chi_{S}(T)$. Initially
$H_{C}(T)$ decreases with temperature due to the unblocking of the
small particles (see Fig.~\ref{hcxs}). Then, thermal fluctuations
leads to a decrease of $\chi_{S}$ and a consequent increase of
$H_{C}$ with $T$ until the large particles start to unblock and
$H_{C}(T)$ decreases again. In the other sample, unblocking occurs
more smoothly in whole temperature range, due to the relative
proximity of the $\left<T_{B1}\right>$ and $\left<T_{B2}\right>$,
and $H_{C}(T)$ present the expected decrease with $T$ (see
Fig.~\ref{hcf}(a)).

\begin{figure}[htbp]
\includegraphics[width=6.0cm]{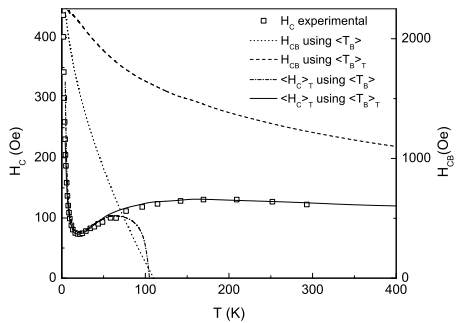}
\caption{\label{hcantf}Coercive field vs. temperature of the
Cu$_{90}$Co$_{10}$ sample: experimental (open squares), calculated
$\left<H_{C}\right>_{T}$ (solid and dotted-dashed lines), and
$H_{CB}$ (dashed and dotted lines). For detail, see text.}
\end{figure}

The above description is particulary adequate for systems that
present considerable interactions and deviations of $H_{C}(T)$
around near $\left<T_{B}\right>$. This was more evident in the
Cu$_{90}$Co$_{10}$ sample due to its unusual behavior. In this
respect we present a comparison of different possible scenarios to
explain the experimental data (see Fig.~ \ref{hcantf}). Without
taking into account the contribution (i) of superparamagnetic
susceptibility of unblocked particles we obtain the curves
indicated by the dotted and dashed lines.

The dotted is the standard and widely used Eqs.~(\ref{Hc}) and
(\ref{tbme}), that works for well isolated and narrow
distributions of sizes.\cite{Bean} The dashed line is determined
by considering the temperature dependent average blocking
temperature (Eqs.~(\ref{Hc}) and (\ref{tbt})) and shows clearly
the need to include the superparamagnetic correction. By
considering further this correction with Eq.~(\ref{tbme}) and
(\ref{hcm}) the dashed-dotted line is obtained with an excellent
agreement at low temperatures, where most of the particles are
still blocked. When we take into account Eqs.~(\ref{tbt}) and
(\ref{hcm}), \textit{i.e.}, considering the temperature dependence
of the average blocking temperature we obtain finally the solid
line describing well the results in the whole temperature range.

\section{ Conclusion}
We present a generalized model for the description of the thermal
dependence of $H_{C}$ of granular magnetic systems. With this
model we describe successfully the temperature dependence of the
coercive field of Co-Cu samples. The contribution of
superparamagnetic particles and the use of a temperature dependent
average blocking temperature was shown to be important to describe
the coercive field in a wide temperature range. The interactions
among the particles were considered through an empirical
multiplying factor $\gamma$ as well as an effective size
distribution. we believe that with this procedure most of the fine
magnetic particle systems may be well described.

\acknowledgments  This work was supported by Instituto de
Nanoci\^encias - Institutos do Mil\^enio - CNPq, FUJB, CAPES and
FAPERJ. The authors thank the project PRONEX/FINEP and Dr. R. S.
de Biasi for helpful discussions.


\begin{thebibliography}{99}
\bibitem{Neel} L. N\'eel, Ann. Geophysique \textbf{5}, 99 (1949).
\bibitem{Fiorani} J. L. Dormann, D. Fiorani and E. Tronc, Adv. Chem. Phys. \textbf{98}, 283 (1997).
\bibitem{Bean} C. Bean and J. D. Livingston, J. Appl. Phys. \textbf{30}, 120S (1959).
\bibitem{Fonseca} F. C. Fonseca, G. F. Goya, R. F. Jardim, R. Muccillo, N. L. V. Carreno, E. Longo and E. R. Leite, Phys. Rev. B \textbf{66},104406 (2002).
\bibitem{Batlle} X. Batlle, M. Garcia del Muro, J.Tejada, H. Pfeiffer, P. Gornert and E. Sinn, J. Appl. Phys. \textbf{74}, 3333 (1993).
\bibitem{Kechakros} D. Kechrakos and K. N. Trohidou, Phys. Rev. B \textbf{58}, 12169 (1998).
\bibitem{Kneller} E. F. Kneller and F. E. Luborsky, J. Appl. Phys. \textbf{34}, 656 (1963).
\bibitem{Pfeiffer} H. Pfeiffer, Phys. Stat. Sol. (a) \textbf{118}, 295 (1990).
\bibitem{Vavassori} P. Vavassori, E. Angeli, D. Bisero, F. Spizzo, and F. Ronconi, Appl. Phys. Lett. \textbf{79}, 2225 (2001).
\bibitem{Nunes} W. C. Nunes, M. A. Novak, M. Knobel e and A. Hernando, J. Magn. Magn. Mater. \textbf{226-230}, 1856 (2001).
\bibitem{Ferrarin} E. F. Ferrari, W. C. Nunes, and M. A. Novak, J. Appl. Phys. \textbf{86}, 3010 (1999).
\bibitem{Peleg} N. Peleg, S. Shtrikman, G. Gorodetsky, and I. Felner, J. Magn. Magn. Mater. \textbf{191}, 349 (1999).
\bibitem{Chantrell} R. W. Chantrell, M. El-Hilo, And K. O'Grady, IEEE Trans. Magn. \textbf{27}, 3570 (1991).
\bibitem{Folly} W. S. D. Folly and R. S. de Biasi, Braz. J. Phys. \textbf{31} (3), 398 (2001).
\bibitem{Kliava} Janis Kliava, and Ren\'e Berger, J. Magn. Magn. Mater. \textbf{205}, 328 (1999).
\bibitem{Allia} P. Allia, M. Knobel, P. Tiberto and F. Vinai, Phys. Rev. B. \textbf{52}, 15398 (1995).
\bibitem{Wecker} J. Wecker, R. von Helmolt, L. Schultz, and K. Samwer, Appl. Phys. Lett. \textbf{62}, 1985 (1993).
\bibitem{Allian} P. Allia, M. Coisson, P. Tiberto, F. Vinai, M. Knobel, M. A. Novak, and W. C. Nunes, Phys. Rev. B \textbf{64}, 144420 (2001).
\bibitem{Stoner} E. C. Stoner and E. P. Wohlfarth, Philos. Trans. R. Soc. London \textbf{A240}, 599 (1948).
\bibitem{elhilo} M. El-Hilo, K. O'Grady and R. W. Chantrell, J. Magn. Magn. Mat., 114, 295 (1992)
\bibitem{mblanco} M. Blanco-Mantecón and K. O'Grady J. Magn. Magn. Mat., 203, 50 (1999)
\bibitem{Ferrari} E. F. Ferrari, F. C. S. da Silva, and M. Knobel, Phys. Rev. B. \textbf{56}, 6086 (1997).
\bibitem{Luis} F. Luis, J. M. Torres, L. M. Garcia, J. Bartolome, J. Stankiewicz, F. Petroff, F. Fettar, J. L. Maurice, and A. Vaures, Phys. Rev. B \textbf{65}, 94409 (2002).
\bibitem{Panissod} P. Panissod, M. Malinowska, E. Jedryka, M. Wojcik, S. Nadolski, M. Knobel, and J. E. Schmidt, Phys. Rev. B \textbf{63}, 014408 (2000).
\bibitem{GarciaPrieto} A. Garc\'ia Prieto, M. L. Fdez-Gubieda, C. Meneghini, A. Garc\'ia-Arribas, and S. Mobilio, Phys. Rev. B \textbf{67}, 224415 (2003).
\bibitem{Lopez} A. L\'opez, F. J. Lázaro, R. von Helmolt, J. L. García-Palacios, J. Wecker, and H. Cerva, J. Magn. Magn. Mater. \textbf{187}, 221 (1998).
\bibitem{Wang} W. Wang, F. Zhu, W. Lai, J. Wang, G. Yang, J. Zhu, and Z. Zhang, J. Appl. Phys. \textbf{32}, 1990 (1999).
\bibitem{Denardin} J. C. Denardin, A. L. Brandl, M. Knobel, P. Panissod, A. B. Pakhomov, H. Liu and X. X. Zhang, Phys. Rev. B \textbf{65}, 064422(2002).
\bibitem{Garciadel} M. Garcia del Muro, X. Batlle and A. Labarta, J. Magn. Magn. Mater. \textbf{221}, 26 (2000).
\bibitem{Respaud} M. Respaud, J. M. Broto, H. Rakoto, A. R. Fert, L. Thomas, B. Barbara, M. Verelst, E. Snoeck, P. Lecante, A. Mosset, J. Osuna, T. O. Ely, C. Amiens and B. Chaudret, Phys. Rev. B \textbf{57}, 2925 (1998).
\bibitem{hutten} A. H\"utten and G. Thomas, Ultramicroscopy \textbf{52}, 581 (1993).
\bibitem{yu} R. H. Yu et al., J. Appl. Phys. \textbf{79(4)}, 1979 (1996).

\end{thebibliography}
\end{document}